\documentstyle[12pt]{article}
\begin{document}

\begin{center}
{\Large \bf Black hole motion in Euclidean space as \linebreak a diffusion process} \\
\vskip .5cm
K. Ropotenko\\
\centerline{\it State Service for Special Communication and}
\centerline{\it Information Protection of Ukraine,} \centerline{\it
5/7 Patorzhynskoho str., Kyiv, 01034, Ukraine}
\bigskip
\verb"ropotenko@ukr.net"

\end{center}
\bigskip\bigskip
\begin{abstract}
A diffusion equation for a black hole is derived from the
Bunster-Carlip equations. Its solution has the standard form of a
Gaussian distribution. The second moment of the distribution
determines the quantum of black hole area. The entropy of diffusion
process is the same, apart from the logarithmic corrections, as the
Bekenstein-Hawking entropy.
\end{abstract}
\bigskip\bigskip

Bunster (Teitelboim) and Carlip showed \cite{car} that the wave
function of a black hole with the Arnowitt-Deser- Misner (ADM) mass
$M$ and area $A$ evolves according to the Schr\"{o}dinger-type
equations
\begin{equation}
\label{eq1} \frac{\hbar}{i}\frac{\partial \psi}{\partial t}+M\psi=0,
\end{equation}
\begin{equation}
\label{eq2} \frac{\hbar}{i}\frac{\partial \psi}{\partial
\Theta}-\frac{A}{8\pi G}\psi=0,
\end{equation}
where $t$ is the lapse of asymptotic proper time at spatial infinity
and $\Theta$ is the lapse of the hyperbolic angle at the horizon.
Under Euclidean continuation $\Theta$ transforms to an angle
variable. As a result, as pointed out in \cite{car}, $A/8\pi G$ and
$\Theta$ become conjugate exactly like $M$ and $t$. This means that
the area is the operator-valued quantity. It was shown in \cite{ro}
that $A/8\pi G$ can be interpreted as the $z$ component of an
internal angular momentum of a black hole $L_z$. Indeed, in the
semiclassical approximation
\begin{equation}
\label{act3} \psi=a\exp \left(\frac{i}{\hbar}I\right),
\end{equation}
where $I$ is the action of a black hole. Substituting this in
(\ref{eq2}) we obtain
\begin{equation}
\label{act4} \psi\frac{\partial I}{\partial \Theta}=\frac{A}{8\pi
G}\psi;
\end{equation}
the slowly varying amplitude $a$ need not be differentiated. Under
Euclidean continuation $\Theta_{\rm E}=i\Theta$ and $I_{\rm E}=iI$,
\begin{equation}
\label{act5} \psi \frac{\partial I_{\rm E}}{\partial \Theta_{\rm E}
}=\frac{A}{8\pi G}\psi.
\end{equation}
The derivative $\partial I_{\rm E}/\partial \Theta_{\rm E}$ is just
a generalized momentum corresponding to the angle of rotation about
one of the axes (say, the $z^{\rm{th}}$) for a mechanical system.
Therefore the operator $A/8\pi G$ is what corresponds in quantum
mechanics to the $z$ component of angular momentum $L_z$.
Quantization of $L_z$ gives the equidistant area spectrum of a black
hole
\begin{equation}
\label{eq6} A_m=\Delta A \cdot m,\quad m=0,1,2,...
\end{equation}
with the area quantum
\begin{equation}
\label{eq7} \Delta A = 8\pi l_{\rm P}^{2}.
\end{equation}
Medved \cite{med} found this value immediately from the
Bunster-Carlip action \cite{car}. Note that Bekenstein \cite{bek}
was the first to determine the quantum of area. Later on, the value
(\ref{eq7}) was obtained using different approaches and techniques
\cite{an}.

In this note I derive a diffusion equation for a black hole from the
Bunster-Carlip equations and show that the black hole motion in
Euclidean space exhibits a diffusion process. Moreover I find that
the entropy of the process is the same, apart from the logarithmic
corrections, as the Bekenstein-Hawking entropy.

I begin with the Bunster-Carlip equations (\ref{eq1}) and
(\ref{eq2}). Analytically continuing  $\Theta$ and $t$ to the real
values of $\Theta_{\rm E}=i\Theta$ and $t_{\rm E}=it$ we obtain
\begin{equation}
\label{eq8} \hbar \frac{\partial \psi}{\partial t_{\rm E}}+M \psi=0,
\end{equation}
\begin{equation}
\label{eq9} \hbar \frac{\partial \psi}{\partial \Theta_{\rm E}}-
\frac{A}{8\pi G}\psi=0.
\end{equation}
Taking the complex conjugate of equation (\ref{eq6}) we get
\begin{equation}
\label{eq10} \hbar \frac{\partial \psi^{\ast}}{\partial \Theta_{\rm
E}}- \frac{A}{8\pi G}\psi^{\ast}=0.
\end{equation}
Multiplying (\ref{eq9}) by $\psi^{\ast}$ and (\ref{eq10}) by $\psi$
and then adding, we find
\begin{equation}
\label{eq11} \hbar \left(\psi^{\ast} \frac{\partial \psi}{\partial
\Theta_{\rm E}}+\psi \frac{\partial \psi^{\ast}}{\partial
\Theta_{\rm E}}\right)= \left(\psi^{\ast} \frac{A}{8\pi G}\psi+\psi
\frac{A}{8\pi G}\psi^{\ast}\right).
\end{equation}
In the spherical polar coordinates, i.e. in terms of $\Theta_{\rm
E}$, the $z$ component of the internal angular momentum $A/8\pi G$
is not a product of operators. Therefore
\begin{equation}
\label{eq12} \hbar \frac{\partial (\psi\psi^{\ast})}{\partial
\Theta_{\rm E}}=\frac{A}{8\pi G}(\psi\psi^{\ast})
\end{equation}
or
\begin{equation}
\label{eq13} \hbar \frac{\partial \rho}{\partial \Theta_{\rm
E}}=\frac{A}{8\pi G}\rho,
\end{equation}
where $\rho=|\psi(t_{\rm E}$, $\Theta_{\rm E})|^{2}$ is the
probability density of finding the black hole at point ($t_{\rm E}$,
$\Theta_{\rm E}$) in the Euclidean manifold. On the other hand, for
a Schwarzschild black hole
\begin{equation}
\label{eq14} \frac{A}{8\pi G}=2GM^{2}.
\end{equation}
Since, according to (\ref{eq8}),
\begin{equation}
\label{eq15} M^{2}\equiv \hbar^{2}\frac{\partial^{2}}{\partial
t_{\rm E}^{2}},
\end{equation}
the equation (\ref{eq13}) reads
\begin{equation}
\label{eq16} \frac{\partial \rho}{\partial \Theta_{\rm
E}}=D\frac{\partial^{2}\rho}{\partial t_{\rm E}^{2}}.
\end{equation}
This is an one-dimensional diffusion equation in the temporal
$\Theta_{\rm E}$ and spatial $t_{\rm E}$ coordinates with the
diffusion coefficient $D=2G\hbar$. The solution to the equation is
\begin{equation}
\label{eq17} \rho=\frac{1}{\sqrt{4\pi D \Theta_{\rm E}}}e^{-t_{\rm
E}^{2}/(4D\Theta_{\rm E})},
\end{equation}
which is normalized such that
\begin{equation}
\label{eq18} \int_{-\infty}^{+\infty}\rho(t_{\rm E},\Theta_{\rm
E})dt_{\rm E}=1.
\end{equation}
It follows that the black hole spreads out with increasing "time"
$\Theta_{\rm E}$. The mean square value of $t_{\rm E}$ is given by
\begin{equation}
\label{eq19} \langle t_{\rm
E}^{2}\rangle=\int_{-\infty}^{+\infty}t_{\rm E}^{2}\rho(t_{\rm
E},\Theta_{\rm E})dt_{\rm E}=2D\Theta_{\rm E}.
\end{equation}
This result shows that the width of distribution increases as
$\Theta_{\rm E}^{1/2}$, which is a general characteristic of
diffusion and random walk problems in one dimension. Since
regularity of the Euclidean manifold at the horizon imposes the
fixed Euclidean angle $\Theta_{\rm E}=2\pi$, we get
\begin{equation}
\label{eq20} \langle t_{\rm E}^{2}\rangle=8\pi l_{\rm P}^{2}.
\end{equation}
This value coincides with (\ref{eq7}). But (\ref{eq7}) is a result
of a true quantum-mechanical quantization of area; it arises due to
the periodicity of $\Theta_{\rm E}$. In contrast, (\ref{eq20}) can
be viewed as a result of discreteness of $\Theta_{\rm E}$ in a
random walk model. In the model, $\langle t_{\rm E}^{2}\rangle=N
l_0^{2}$, where $l_0$ is the length of each step of a random walk
and $N$ is the number of steps. The quantum of area $8\pi l_{\rm
P}^{2}$ arises if we let $2\pi$ be the duration of a step; then
$\langle t_{\rm E}^{2}\rangle =(\Theta_{\rm E}/2\pi)l_0^{2}$ which
being combined with (\ref{eq19}) gives $l_0^{2} = 8\pi l_{\rm
P}^{2}$. By abuse of language, we call $8\pi l_{\rm P}^{2}$ in
(\ref{eq20}) the quantum of area. Therefore the motion of a black
hole in Euclidean space exhibits a diffusion process. Thus the
elementary act which changes the size of a black hole is the gain or
loss by it, of one quantum of area $8\pi l_{\rm P}^{2}$ during one
period $\Theta_{\rm E}=2\pi$.

Analytical continuation to the cyclic imaginary time $\Theta_{\rm
E}$ means that we deal with a quantum system at a finite
temperature.  In this case, the system is described not by the
probability density $\rho (x)$ but by the density matrix $\rho
(x,x^{\prime};\beta)$, where $x$ is a spatial coordinate and
$\beta\equiv T^{-1}$ is the inverse temperature. The partition
function $Z(\beta)$ is the trace of the density matrix
\begin{equation}
\label{eq21} Z(\beta)=\int \rho (x,x;\beta) dx.
\end{equation}
In (\ref{eq16}), $t_{\rm E}$ plays the role of the spatial
coordinate. The temporal coordinate $\Theta_{\rm E}= k t_{\rm E}$,
where $k=1/4GM$ is the surface gravity. Since $\Theta_{\rm E}$ has
the interpretation as an angular coordinate with periodicity $2\pi$,
$t_{\rm E}$ itself has then periodicity $8\pi GM$ which, when set
equal to $\hbar/T_{\rm H}$, gives the Hawking temperature $T_{\rm
H}$. Since $t_{\rm E}\sim t_{\rm E}+8\pi GM$, we have
\begin{equation}
\label{eq22} \rho (t_{\rm E},t_{\rm E};\beta)=\frac{1}{\sqrt{4\pi D
\Theta_{\rm E}}}e^{-\beta^{2}/(4D\Theta_{\rm E})}
\end{equation}
Therefore
\begin{equation}
\label{eq23} Z(\beta)=\frac{\beta}{\sqrt{4\pi D \Theta_{\rm
E}}}e^{-\beta^{2}/(4D\Theta_{\rm E})}.
\end{equation}
The internal energy is given by
\begin{equation}
\label{eq24} E=-\frac{\partial \ln Z(\beta)}{\partial
\beta}=M-\frac{1}{8\pi GM}.
\end{equation}
It is the same, apart from a term of order of the Hawking
temperature, as the ADM mass of a black hole. Finally, the entropy
is given by
\begin{equation}
\label{eq25} S=\ln Z +\beta E =\frac{A}{4l_{\rm
P}^{2}}+\frac{1}{2}\ln \left(\frac{A}{4l_{\rm
P}^{2}}\right)+\ln\left(\frac{1}{e\sqrt{\pi}}\right).
\end{equation}
It is the same, apart from the terms $\mathcal{O }$$(\ln (A/4l_{\rm
P}^{2}))$, as the Bekenstein-Hawking entropy $S_{\rm BH}=A/4l_{\rm
P}^{2}$.

$\linebreak$

I thank an anonymous referee for drawing my attention to a mistake
in deriving the diffusion equation in the previous version of the
paper.

\end{document}